\documentclass[amssymb,prb,twocolumn,superscriptaddress]{revtex4-1}

\usepackage{bm}
\usepackage{epsfig}
\usepackage{graphicx}
\usepackage{amsmath}

\usepackage[%
colorlinks=true,
urlcolor=blue,
linkcolor=blue,
citecolor=blue
]{hyperref}

\renewcommand{\Re}{{\mathrm{Re}}\, }
\renewcommand{\Im}{{\mathrm{Im}}\, }

\newcommand{\ket}[1]{|#1\big>}
\newcommand{\ra}{\rightarrow}

\newcommand{\comment}[1]{}
\newcommand{\bk}{{\bf k}}
\newcommand{\bp}{{\bf p}}
\newcommand{\bkp}{{\bk+\bp}}
\newcommand{\bd}{{\bf d}}
\newcommand{\br}{{\bf r}}

\newcommand{\Tr}{{\rm Tr}}

\begin{document}

\title{Local and global patterns in quasiparticle interference: a reduced response function approach}

\author{Dan-Bo Zhang}
\affiliation{Guangdong Provincial Key Laboratory of Quantum Engineering and Quantum Materials, GPETR Center for Quantum Precision Measurement, SPTE, South China Normal University, Guangzhou 510006, China}

\author{Qiang Han}
\email{hanqiang@ruc.edu.cn}
\affiliation{Department of Physics, Renmin University of China,
Beijing, China}

\author{Z. D. Wang}
\email{zwang@hku.hk}
\affiliation{Department of Physics and Center of Theoretical and
Computational Physics, The University of Hong Kong, Pokfulam Road,
Hong Kong, China}
\affiliation{Guangdong Provincial Key Laboratory of Quantum Engineering and Quantum Materials, GPETR Center for Quantum Precision Measurement,  SPTE, South China Normal University, Guangzhou 510006, China}
\date{\today}
\begin{abstract}
A physical system exposes to us in a real space, while its description often refers to its reciprocal momentum space. A connection between them can be established by
exploring patterns of quasiparticles interference (QPI), which is experimentally accessible by Fourier transformation of the scanning tunneling spectroscopy~(FT-STS). We here investigate how local and global features of QPI patterns are related to the geometry and topology of electronic structure in the considered physical system.  A reduced response function (RRF) approach is developed that can analyze QPI patterns with clear physical pictures. It is justified that the generalized joint density of states, which is the imaginary part of RRF, for studying QPI. Moreover, we reveal that global patterns of QPI may be indicators of topological numbers for gapless systems, and demonstrate that robustness of such indicators against distractive local features of QPI for topological materials with complicated band structures.

\end{abstract}

\maketitle
%
\section{Introduction}
Interference of quasiparticles in the presence of impurities
leads to a modulation of the local density of states (LDOS). The LDOS is accessible experimentally with scanning tunneling spectroscopy (STS)\cite{crommie_93,hasegawa_93,sprunger_97}, and the pattern of modulation is further extracted by its Fourier transformation. There are local and global patterns of QPI that depend on the band structure of the underlying physical system. Concisely, local patterns of QPI depend on the geometry of the dispersion \cite{capriotti_03}, and are further modified (suppressed or enhanced) by the internal structure of quasiparticles~\cite{wang_03,hanaguri_09}. Global patterns of QPI, on the other hand, depend on the topology of band structure, e.g., a global structure of spin-momentum locking~\cite{guo_10}. Consequently, we can use QPI to infer the geometry and topology of the band structure, rendering QPI and FT-STS as facilities bridging the $\bf{r}$-space
observations with the $\bf{k}$-space description of the underlying system.

There have been intensive studies of QPI for quantum materials~\cite{avraham_18}, including metals~\cite{crommie_93,sprunger_97}, superconductors~\cite{hoffman_02,capriotti_03,wang_03,pereg-barnea_03,hanaguri_09,lee_09,zhang_13}, graphene~\cite{rutter_07,pereg-barnea_08,brihuega_08,dombrowski_17,jolie_18}, surface states of topological insulators~\cite{zhou_09,lee_09_TI,beidenkopf_11,kohsaka_15,kohsaka_17}, Weyl semimetals~\cite{batabyal_16,inoue_16,mitchell_16,zheng_18}, and nonsymmorphic materials~\cite{topp_17,queiroz_18,zhu_18}. The connection between QPI and $\bk$-space spectral information is established through the joint density of states (JDOS) and its generalization (GJDOS) that takes internal structures of quasiparticles into consideration. The JDOS approach has been proposed at an earlier stage to explain hot spots in patterns of QPI for d-wave superconductor~\cite{hoffman_02,capriotti_03,wang_03,mcelroy_03}, and has been applied for gapless topological systems in terms of spin-selective scattering probability (SSP) recently~\cite{beidenkopf_11,inoue_16}. A more rigorous treatment, however, should refer to Fourier transform of LDOS (FT-LDOS). Ref.~\cite{derry_15,kohsaka_17} point out that GJDOS may give some false features, and is not appliable in general. Nevertheless, as a convenient tool GJDOS allows us to intuitively analyze and understand QPI patterns from band structures directly. It is desirable for a clarification of the applicability of GJDOS. Moreover, a systematic treatment that can relate local and global patterns of QPI to the geometry and topology of the band structure is still awaited.

In this paper, we adopt a reduced response function~(RRF) approach for analyzing QPI patterns under different scattering and probing channels. The RRF includes GJDOS as its imaginary component, and is a faithful encoding of the QPI information.  We will prove that GJDOS shares the same singularities with FT-LDOS, with few exceptions that can be excluded at first hand. Moreover, we reveal the existence of higher-degree singularities, which can give rise to more significant features (hot spots) in QPI. With the justified GJDOS, we derive its analytical expression that clearly shows how its singularities and coherent factors jointly determine the QPI patterns. Based on this expression, we propose indicators of topological numbers in ideal topological systems. We then testify and demonstrate the robustness of such indicators of topological numbers against distracting local features in QPI arising from complicated geometry of the band structure, by exploring the QPI patterns in several representative topological materials, including topological insulators $\text{Bi}_2\text{Te}_3$ and BiTeI, as well as the graphene family.

The paper is organized as follows. We first propose the reduced response function and derive its singular behavior in Sec.~\eqref{sec:rrf}. Then, we give an analytical expression for GJDOS and apply it for analyzing global patterns in Sec.~\eqref{sec:global_patterns}. We then numerally study QPI for topological materials in Sec.~\eqref{sec:qpi-materials}, including surface states of topological insulators and the graphene family.

\section{Reduced Response function for quasiparticle interference}
\label{sec:rrf}
\subsection{The reduced response function}
For simplicity without loss of generality, we restrict our discusses of physical systems described by two-band Hamiltonians $\mathcal{H}(\bk)=E_0(\bk)+\mathbf{d}(\bk)\cdot\boldsymbol{\sigma}$. Here $\boldsymbol{\sigma}=(\sigma_x,\sigma_y,\sigma_z)$ denotes a  vector of Pauli matrix for a spin (or a pseudospin), which can represent systems with two internal degrees of freedom, including spin-half, sublattice, or particle-hole.  Impurity is described as $V(r)=\sum_{\beta}V^\beta(r)\sigma_\beta$, where $\beta=0,x,y,z$ stands for scattering channels. In the presence of an impurity, interference between scattered-in and out quasiparticles leads to a perturbation to the local density of states. The local density is related to the Green function, \begin{equation}
n_{\alpha}(\br,\omega)=-\frac{1}{\pi}\Im\{\Tr[\sigma_\alpha G(\br,\br,\omega)]\},
\end{equation}
where $\sigma_\alpha$ ($\alpha=0,x,y,z$) represents probe channels~\cite{guo_10}. The Green function in $\bk$-space can be written as
\begin{equation}
G(\bk',\bk,\omega)=G_0(\bk,\omega)\delta_{\bk\bk'}+G_0(\bk',\omega)T_{\bk'\bk}(\omega) G_0(\bk,\omega),
\end{equation}
where $G_0(\bk,\omega)=[\omega-\mathcal{H}(\bk)+i0^+]^{-1}$ is
the unperturbed Green function with positive infinitesimal $0^+$. For a $\sigma_\beta$ impurity and under Born approximation, the T-matrix takes the form $T_{\bk'\bk}(\omega)=V^{\beta}_{\bk'\bk}\sigma_\beta$, where $V^{\beta}_{\bk',\bk}$ is a Fourier transform of the potential strength for scattering channel $\sigma_\beta$.
The scattering potential may be independent of $\bk'$ and $\bk$, e.g., for a impurity with a delta potential $\delta(\br)$. Or it can depend on $\bk'-\bk$~\cite{capriotti_03}.
We mention that in general it can depend on both $\bkp$ and $\bk$, e.g., for spin-orbit scattering the scattering potential is $V_{\bkp,\bk}=V_0\{1+ic[(\bkp)\times\bk]\cdot\boldsymbol{\sigma}\}$~\cite{lee_09_TI,kohsaka_17}.

The perturbed local density for probing channel $\sigma_\alpha$ and scattering channel $\sigma_\beta$ turns to be
\begin{equation}
\delta n_{\alpha\beta}(\br,\omega)=-\frac{1}{\pi}\Im\{\Tr[\sum_{\bk'\bk}\sigma_\alpha G_0(\bk',\omega)V^{\beta}_{\bk',\bk}\sigma_\beta G_0(\bk,\omega)]\}.
\end{equation}

QPI patterns can be captured by a Fourier transformation of the perturbed local density,
\begin{eqnarray}\label{def:ldos_q}
\delta n_{\alpha\beta}(\bp,\omega)&\equiv& \sum_{\br}  e^{-i\bp\cdot\br} \delta n_{\alpha\beta}(\br,\omega)\nonumber \\
&=&-{1\over 2\pi i}
[\Lambda_{\alpha\beta}(\bp,\omega)-\Lambda^*_{\alpha\beta}(-\bp,\omega)],
\end{eqnarray}
where the response function reads
\begin{equation}\label{lam1}
\Lambda_{\alpha\beta}(\bp,\omega) =\sum_\bk V^{\beta}_{\bkp,\bk}\Tr\left[\sigma_\alpha G_0(\bk+\bp,\omega)\sigma_\beta G_0(\bk,\omega)\right].
\end{equation}
In the presence of centrosymmetry, $\Lambda_{\alpha\beta}(\bp,\omega)=\Lambda_{\alpha\beta}(-\bp,\omega)$, Eq.~\eqref{def:ldos_q} reduces to $\delta n_{\alpha\beta}(\bp,\omega)=-\frac{1}{\pi}\Im[\Lambda_{\alpha\beta}(\bp,\omega)]$. We consider the centrosymmetry in this paper since it is applicable for many QPI, while the case
$\Lambda_{\alpha\beta}(\bp,\omega)\neq\Lambda_{\alpha\beta}(-\bp,\omega)$ is left for further investigation.

To establish a connection between FT-LDOS and GJDOS, we first decompose the response function $\Lambda_{\alpha\beta}(\bp,\omega)$. Note $H(\bk)\ket{\psi_{\bk}^{\pm}}=E_{\bk}^\pm\ket{\psi_{\bk}^{\pm}}$, with eigenstates
$\psi_{\bk}^{\pm}(\mathbf{r})\sim u_{\bk,\pm}\text{exp}(i\bk\cdot\mathbf{r})$ and eigenvalues $E_{\bk}^\pm=E_0\pm d_\bk$. Here we write $\mathbf{d(\bf{k})}=d_\bk\hat{\bf{d}}_\bk$ with $d_\bk=|\mathbf{d(\bf{k})}|$ .  Then  $\Lambda_{\alpha\beta}$ can be explicitly written as,
\begin{eqnarray}
\Lambda_{\alpha\beta}(\bp,\omega)=\sum_{\bk ss'}\frac{P_{\alpha\beta}F^{ss'}_{\alpha\beta}(\bkp,\bk,\omega)}{(\omega_\bkp-s'd_\bkp+i0^+)(\omega_\bk-sd_\bk+i0^+)},\nonumber \\
\end{eqnarray}
where
\begin{equation}
P_{\alpha\beta}F^{ss'}_{\alpha\beta}=\Tr\left[ V^{\beta}_{\bkp,\bk}(u_{\bk+\bp,s'}^{\dagger} \sigma_\alpha u_{\bk,s}) (u_{\bk,s}^{\dagger}\sigma_\beta u_{\bk+\bp,s'})\right].
\end{equation}
Here we have briefly written $\omega_\bk=\omega-E_0(\bk)$. $F^{ss'}_{\alpha\beta}$ is the spin coherent factor and we set as a real function (see Sec.~\eqref{sec:analysis_F} for more details), and the complex component has been put in $P_{\alpha\beta}$, which is momentum independent if we assume  $V^{\beta}_{\bkp,\bk}=e^{i\phi_\beta}|V^{\beta}_{\bkp,\bk}|$, e.g., the phase shift in the scattering is  independent of $\bp$.
The dominator in $\Lambda_{\alpha\beta}(\bp,\omega)$ contributes to singularities and the numerator accounts for further enhancements or depressions, and a combination of both can explain hot-spot features in QPI. Introducing
\begin{eqnarray}
&&A_s(\bk,\omega)\equiv-{1\over \pi}
{\Im} [\frac{1}{\omega_\bk-sd_\bk+i0^+}] =\delta(\omega_\bk-sd_\bk) \nonumber \\
&&B_s(\bk,\omega)\equiv-{1\over \pi}
{\Re}[\frac{1}{\omega_\bk-sd_\bk+i0^+}], \nonumber
\end{eqnarray}
we can write FT-LDOS (omitting a factor of $-\frac{1}{\pi}$),
\begin{eqnarray}\label{eq:RF-AA}
&&\delta n_{\alpha\beta}(\bp,\omega)\nonumber \\
&=&\Re[P_{\alpha\beta}]\sum_{\bk ss'} F^{ss'}_{\alpha\beta}[A_s(\bkp,\omega)B_{s'}(\bk,\omega)+A \leftrightarrow B]\nonumber \\
&+& \Im[P_{\alpha\beta}]\sum_{\bk ss'} F^{ss'}_{\alpha\beta}[A_s(\bkp,\omega)A_{s'}(\bk,\omega)+A \leftrightarrow B]. \nonumber \\
\end{eqnarray}

Since an  autocorrelation of $AA$-type dominates typically a $BB$-type one in the second line of Eq.~\eqref{eq:RF-AA}, we chose only the $AA$ term which corresponds to the joint density of states. As for an approximate yet faithful encoding of $\delta n_{\alpha\beta}(\bp,\omega)$, we propose a so-called reduced response function $\mathcal{R}_{\alpha\beta}(\bp,\omega)$
\begin{equation} \label{def:RRF}
\mathcal{R}_{\alpha\beta}(\bp,\omega) = \mathcal{S}_{\alpha\beta}(\bp,\omega) + i \mathcal{J}_{\alpha\beta}(\bp,\omega),
\end{equation}
where
\begin{eqnarray}\label{eq:S&J}
\mathcal{S}_{\alpha\beta}(\bp,\omega)&=&\sum_{\bk ss'}F^{ss'}_{\alpha\beta}[A_s(\bkp,\omega)B_{s'}(\bk,\omega)+A \leftrightarrow B]\nonumber \\
\mathcal{J}_{\alpha\beta}(\bp,\omega)&=& \sum_{\bk ss'}F^{ss'}_{\alpha\beta}[A_s(\bkp,\omega)A_{s'}(\bk,\omega)].
\end{eqnarray}
Note that $\mathcal{S}$ and $\mathcal{J}$ are related by the Hilbert transformation (or the Kramers-Kronig relation). The generalized joint density of states (GJDOS) $\mathcal{J}$ incorporates spin coherent factor into JDOS, and has been applied to analyze QPI for gapless topological systems in terms of spin-selective scattering probability~\cite{beidenkopf_11,batabyal_16}.
The reduced response function contributes to FT-LDOS $\delta n_{\alpha\beta}(\bp,\omega)$ by a projection onto the direction $P_{\alpha\beta}$ in the complex plane, namely
\begin{equation}
n_{\alpha\beta}(\bp,\omega)\simeq \frac{1}{2}(P_{\alpha\beta}^*\mathcal{R}_{\alpha\beta}(\bp,\omega) +h.c.).
\end{equation}

\subsection{Spin coherent factor} \label{sec:analysis_F}
Now we derive the expression of $P_{\alpha\beta}F^{ss'}_{\alpha\beta}$ and reveal the meaning of the spin coherent factor $F^{ss'}_{\alpha\beta}$.  We exploit a representation of density matrix  $\rho_{\bk,s}\equiv u_{\bk,s}u_{\bk,s}^+=\frac{1}{2}(1+s\hat{\bd}_\bk\cdot\boldsymbol{\sigma})$. Since $\Tr[abc]=\Tr[bca]$, we have
\begin{eqnarray}
&&P_{\alpha\beta}F^{ss'}_{\alpha\beta}(\bkp,\bk,\omega)\nonumber\\
&=&
\nonumber V^{\beta}_{\bkp,\bk} \Tr[\sigma_\alpha\rho_{\bkp,s'}\sigma_\beta \rho_{\bk,s}]\\
&=&\frac{1}{4}V^{\beta}_{\bkp,\bk} \sum_{jl}\Tr[\sigma_\alpha\sigma_\beta+\sigma_j\sigma_\alpha\sigma_l\sigma_\beta\hat{d}^j_\bk\hat{d}^l_\bkp] \nonumber
\end{eqnarray}
It should be reminded that $j,l=x,y,z$ while $\alpha,\beta=0,x,y,z$. Some interesting results can be immediately derived for the summation term denoted as $t^{ss'}_{\alpha\beta}(\bkp,\bk)$.
For $\alpha$ and $\beta$, there are three different cases:
\begin{enumerate}
	\item $\alpha=\beta$. Then $t^{ss'}_{\alpha\alpha}=2(1+ss'\hat{\bd}_\bkp\cdot\hat R_\alpha\hat{\bd}_\bk)$. Here $\hat{R}_\alpha$ is a mirror reflection with the $\alpha=x,y,z$ axis and $R_0$ means no operation.
	\item $\alpha=0,\beta\neq 0$. Then $t^{ss'}_{0\beta}=2iss'L^\beta$, where $(L^x, L^y,L^z)=\hat{\bd}_\bkp\times\hat{\bd}_\bk$. By symmetry we have $t^{ss'}_{\beta0}=t^{ss'}_{0\beta}$.
	\item $\alpha\neq 0,\beta\neq 0,\alpha$. Then $t^{ss'}_{\alpha\beta}=2ss'(\hat{d}^\alpha_\bk\hat{d}^\beta_\bkp+\hat{d}^\beta_\bk\hat{d}^\alpha_\bkp)$.
\end{enumerate}
Here the second case is remarkable. An factor of $i$ appears here. If $V^{\beta}_{\bkp,\bk}$ is a real number, then we have $P_{0\beta}=i$, meaning that the GJDOS should account for quasiparticle interference and thus is expected to give sharp QPI features. However, those features may be depressed, for instance, in backscattering processes, where $t^{ss'}_{0\beta}(-\bk,\bk)=0$.
Nevertheless, GJDOS can be used to explain QPI from scattering between quasiparticles locating at cusps of CCE for time-reversal breaking systems, such as d-wave superconductors~\cite{capriotti_03,wang_03}.

\subsection{Singular behaviors of the reduced repsonse function}
The reduced response function can be written as two parts. The first part corresponds to terms of $AB+iAA$ type and the second part corresponds to $BA+iAA$ type, as written in Eq.~\eqref{def:RRF} and Eq.~\eqref{eq:S&J}. In the following we give a derivation of the first part and  the case of the second part can be obtained similarly. Our result shows that both the real and imaginary parts share the same singularities. To focus on the singular behavior, we omit the factor $F_{\alpha\beta}^{ss'}$ temporarily. The first part of RRF is 
\begin{eqnarray}
&&\sum_{ss'}r^{ss'}_{AB}(\bp,\omega)=\nonumber \\
&&\sum_{\bk ss'}\left[A_s(\bkp,\omega)B_{s'}(\bk,\omega)+A_s(\bkp,\omega)A_{s'}(\bk,\omega)\right] \nonumber
\end{eqnarray}
Using $1/(x+0^+)=P(1/x)-i\pi\delta(x)$, the first part of RRF can be rewritten as
\begin{equation} \label{eq:cal_RF}
\begin{aligned}
\sum_{ss'}r^{ss'}_{AB}(\bp,\omega)&=\sum_{ss'}\int \frac{1}{\omega-E^{s'}_\bkp+i0^+} \delta(\omega-E^s_\bk) dk_xdk_y\\
&=\sum_{ss'}\int_{E^s_\mathbf{k}=\omega} \frac{1}{\omega-E^{s'}_\mathbf{k+p}+i0^+}\frac{1}{|\nabla_\mathbf{k} E^s_\mathbf{k}|}dl.
\end{aligned}
\end{equation}
The second part $\sum_{ss'}r^{ss'}_{BA}(\bp,\omega)$ is obtained by a replacement $\bkp\leftrightarrow\bk$.
\begin{figure}
	\includegraphics[width=0.8\columnwidth]{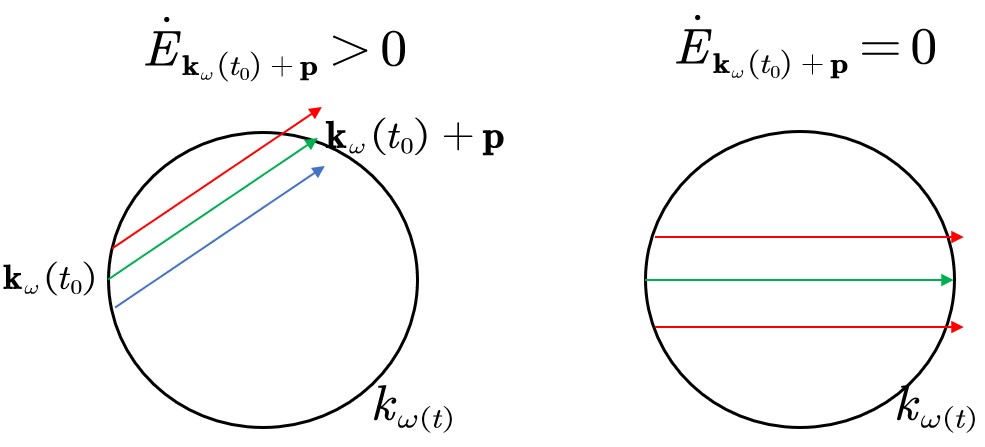}
	\caption{Behavior of $\dot{E}_{\bk_\omega(t_0)+\bp}$. Blue and red lines stand for a small departure of scattered-in quasiparticle from $\bk_\omega(t_0)$ but with the same $\bp$. The left figure shows the general case while the right demonstrates the case of backscattering.}
	\label{fig:backscattering_math}
\end{figure}
The integral is along CCEs of $E^s_\mathbf{k}=\omega$. We may parametrize the $n$-th CCE as $\mathbf{k}^{n,s}_\omega(t)=(f^{n,s}_\omega(t),g^{n,s}_\omega(t))$
with $t\in [0,2\pi)$.  For brevity we would omit the index $n$.   After a tedious derivation 
 we can show that the integral is singular for $\bp$ under a given energy $\omega$ only if there exists $t_0$ that
\begin{eqnarray}\label{eq:singularity}
&&\omega-E^{s'}_{\mathbf{k}^s_\omega(t_0)+\mathbf{p}}=0 \nonumber\\
&&\dot{E}^{s'}_{\mathbf{k}^s_\omega(t_0)+\mathbf{p}}=\left.\frac{dE^{s'}_{\mathbf{k}^s_\omega(t)+\mathbf{p}}}{dt}\right|_{t=t_0}=0.
\end{eqnarray}
The first condition promises that the scattered quasiparticle is on the CCE, and the second one points out that $\bp$ should be specified. An illustration can be seen in Fig.~\eqref{fig:backscattering_math} to show that backscattering meets the above conditions. Moreover, Eq.~\eqref{eq:singularity} servers as a mathematical foundation for exploring singular behavior analytically.

To study the singular behavior, we evaluate $r^{ss'}_{AB}(\mathbf{p}+\bm{\delta},\omega)$ with $|\bm{\delta}|\sim0$. An expression can be found as
\begin{widetext}
	\begin{equation}
	\label{eq:sing_jcurvature}
	r^{ss'}_{AB}(\mathbf{p}+\bm{\delta},\omega) \sim
	- \frac{1}{|\mathbf{v^s}_{\mathbf{k}_0}||\mathbf{v^{s'}}_{\mathbf{k}_0+\mathbf{p}}|}
	\frac{1}{|\bm{\delta}\cdot\bm{\varrho}^{ss'}_{\mathbf{k}_0+\mathbf{p},\mathbf{k}_0}|^{1/2}}
	\left\{
	\begin{aligned}
	& \text{sgn}(\mathbf{n}_{\mathbf{k}_0+\mathbf{p}}\cdot\bm{\varrho}^{ss'}_{\mathbf{k}_0+\mathbf{p},\mathbf{k}_0}), & \bm{\delta}\cdot\bm{\varrho}^{ss'}_{\mathbf{k}_0+\mathbf{p},\mathbf{k}_0} > 0 & \\
	& i, & \bm{\delta}\cdot\bm{\varrho}^{ss'}_{\mathbf{k}_0+\mathbf{p},\mathbf{k}_0} < 0 &
	\end{aligned}
	\right.
	,
	\end{equation}
\end{widetext}
where we have defined the joint curvature,
\begin{equation}\label{j_curvature}
\bm{\varrho}^{ss'}_{\mathbf{k}',\mathbf{k}}=\kappa^{s'}_{\mathbf{k}'}\mathbf{n}^{s'}_{\mathbf{k}'}- \kappa^{s}_{\mathbf{k}}\mathbf{n}^{s}_{\mathbf{k}}.
\end{equation}
Here $\kappa^s_\mathbf{k}$ is the curvature of the CCE $\omega=E^s_\bk$ at $\mathbf{k}$, $\mathbf{n}^s_\mathbf{k}=\mathbf{v}^s_\mathbf{k}/|\mathbf{v}^s_\mathbf{k}|$ the direction vector of
the group velocity, $\mathbf{k}_0$  the real solution of Eq.~\eqref{eq:singularity}. We remark a similar expression can be found for $r^{ss'}_{AB}(\mathbf{p}+\bm{\delta},\omega)$ by an exchange of $\bk_0+\bp \leftrightarrow \bk_0$.

\begin{figure}[ht]
 	\includegraphics[width=0.4\columnwidth]{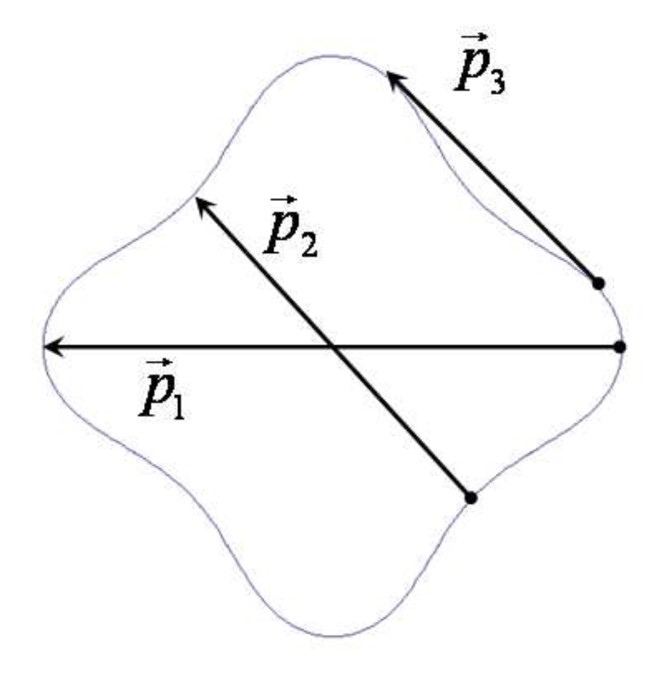}
 	\caption{CCE that changes sign of the curvature for a toy model. $\mathbf{p}_{1,2,3}$ are singularities of $R_{\alpha\beta}(\mathbf{p},\omega)$ because they connect points with their group velocities (anti)parallel. According to Eq.~\eqref{j_curvature}, only the joint curvature of $\mathbf{p}_3$ is zero which leads to the fact that the order of $\mathbf{p}_3$ is higher than those of $\mathbf{p}_{1,2}$.}
 	\label{fig:toy-jont-curvature}
\end{figure}

The expression of singular behavior of Eq.~\eqref{eq:sing_jcurvature}
in terms of the joint curvature provides a clear geometrical tool for analysis: (1) The real and imaginary parts of $R(\mathbf{p},\omega)$ share the same singularities and diverge complementarily in a inverse-square-root manner, unless the joint curvature is zero. (2) If the joint curvature of certain singularity is zero, the exponent of the power-law divergence is  $-2/3$ and can be even higher. 
In Fig.~\eqref{fig:toy-jont-curvature}, we illustrate singularities with nonzero and zero joint curvatures, and they gives divergence of $-\frac{1}{2}$ and at least $-\frac{2}{3}$, separately.

\subsection{Stability of singularities for  finite life-time quasiparticles}
We consider the stability/robustness of singularities of $\mathcal{R}$ when quasiparticles have finite life-time. We argue that singularities are not stable for $\mathcal{S}$ when a length of CCE is approximate flat. We simply take $0^+\rightarrow \eta$, where $\eta$ is a positive finite number accounting for the finite life-time for quasiparticles.

A finite life-time of quasiparticle will
smooth singularities.  Note that $B(\bk,\omega)$ and $A(\bk,\omega)$ exhibit distinct behavior around $\omega=E_\bk$(see Fig.~(3a)). Remarkably, $B(\bk,\omega)$ is zero at  $\omega=E_\bk$ and changes sign at two sides. The real part of RRF, $\mathcal{S}$, is an autocorrelation of $A(\bk,\omega)$ and $A(\bk,\omega)$, has a large weighting at $|\omega-E_\bk|<\eta$. For a large-momentum backscattering in Fig.~(3b), scattered-out quasiparticle always has energy $\omega>E_\bk$, and thus gives a positive contribution to $\mathcal{S}$ . However, when scattering occurs on an approximated flat CCE (compared to $|\bp|$), (see Fig.~(3c)), then scattered-out quasiparticle of both $\omega>E_\bk$ and $\omega<E_\bk$ contribute to $\mathcal{S}$, and those contributions cancel out. The case for near-nesting can be analyzed similarly.  In contrast, the joint density of states $\mathcal{J}$ is an autocorrlation of two $A(\bk,\omega)$, which receive always positive contributions around and mainly from $\omega=E_\bk$. Thus, when QPI pattern should be given by $\mathcal{S}$, $\mathcal{J}$ would give false features for scattering on an approximately flat band. For instance, a hot spot at $\bp=0$ always presents in $\mathcal{J}$ but is absent in $\mathcal{S}$ for $\eta\neq 0$. This can clarify a discrepancy between FT-LDOS and GJDOS for explaining QPI~\cite{derry_15}.

\begin{figure}[ht]
	\includegraphics[width=0.8\columnwidth]{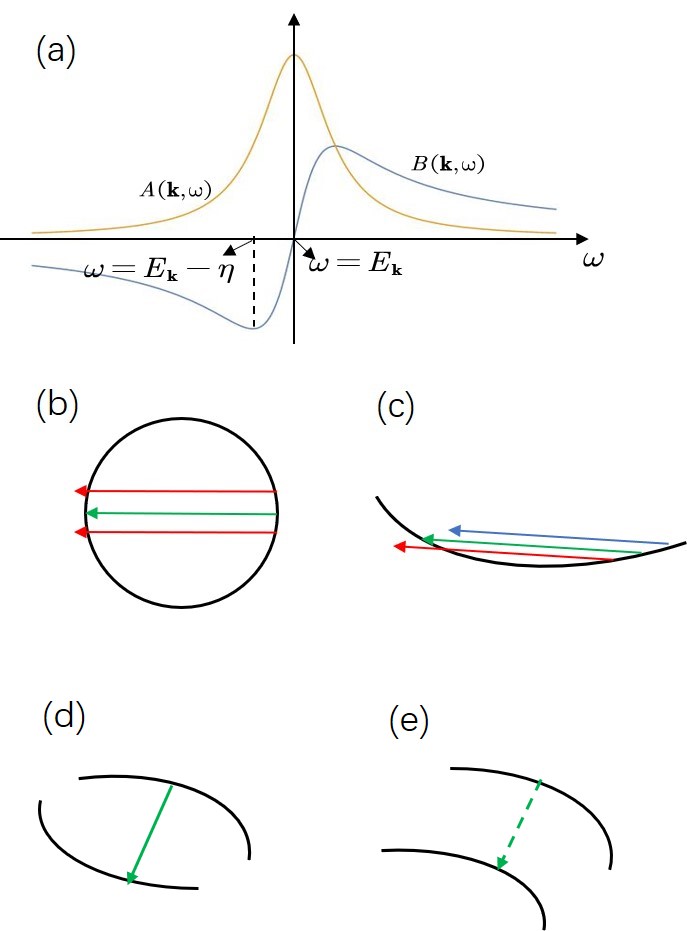}
	\caption{Stability of singularities for the real part of the reduced response function $\mathcal{S}$. (a) Amplitude of $A(\bk,\omega)$ and $B(\bk,\omega)$ with varying $\omega$ under positive finite $\eta$ that models a finite life-time of quasiparticle. (b). Stable singularity. Contributions to $\mathcal{S}$ are all positive (red arrows) under fixed $\bp$ (green arrow). (c). Unstable singularity.  Positive (red arrow) and negative (blue arrow) contributions to $\mathcal{S}$ under fixed $\bp$. (d). Stable singularity for near-nesting scattering. (e). Unstable singularity for near-nesting scattering. }
	\label{fig:S_J_lifetime}
\end{figure}

\section{Global patterns in QPI for ideal topological systems}
\label{sec:global_patterns}
In the above section, we have revealed that $\mathcal{S}$ and $\mathcal{J}$ share same singularities ideally. While in a real application (where quasiparticles have finite life-time), singularities of $\mathcal{S}$ disappear for approximately flat CCE, but are remained in $\mathcal{J}$, leading to a discrepancy. However, those discrepancies correspond to rather specified QPI patterns, such as hot spot or an asterisk-like pattern around $\bp=0$~\cite{kohsaka_15}. Thus, we can take them as exceptions at first hand when using $\mathcal{J}$. After having justified the applicable of $\mathcal{J}$,  we will derive a simple expression of $\mathcal{J}$ that can be used directly for analyzing QPI pattern in a geometrical fashion.  We then use this expression to analyze global patterns in QPI for ideal topological Dirac points with varied topological numbers.

\subsection{Analytical expressions of GJDOS}
Due to the properties of $\delta$-function，
$J^{ss'}_{\alpha\beta}(\bp,\omega)$ can be analytically evaluated, :
\begin{equation}
\begin{aligned}
J_{\alpha\beta}(\bp,\omega) &= \int F^{ss'}_{\alpha\beta}(\bkp,\bk) \frac{\delta(\omega-E^\prime) \delta(\omega-E)}{\left|\frac{\partial (E^{s'}_\mathbf{k+p}, E^s_\mathbf{k})}{\partial (k_x, k_y)}\right| } dEdE^\prime \\
&= \sum_{\mathbf{k}_0ss'}  F^{ss'}_{\alpha\beta}(\bkp,\bk)\left|\frac{\partial (E^{s'}_\mathbf{k+p}, E^s_\mathbf{k})}{\partial (k_x, k_y)} \right|^{-1}_{\mathbf{k}=\mathbf{k}_0}, \\
&= \sum_{\mathbf{k}_0ss'} \frac{F^{ss'}_{\alpha\beta}(\bk_0+\bp,\bk_0)}{\left| \mathbf{v}^{s'}_{\mathbf{k}_0+\bf{p}}\times \mathbf{v}^s_{\mathbf{k}_0}\right|},
\end{aligned} \label{JDOS_exact}
\end{equation}
where the sum is taken over all $\mathbf{k}_0$'s in the real solution set of the following two equations,
\begin{eqnarray}
E^{s'}_{\mathbf{k}_0+\bf{p}}=\omega, \label{CCE1} \\
E^s_{\mathbf{k}_0}=\omega. \label{CCE2}
\end{eqnarray}

The singularity is obtained by setting the dominator
\begin{equation}
\label{jacobi}
\mathbf{v}^{s'}_{\mathbf{k}_0+\bf{p}} \times \mathbf{v}^s_{\mathbf{k}_0} = 0,
\end{equation}
which could be seen as a generalization of Van Hove singularity with regards to joint density of states. The condition $\eqref{jacobi}$ is reached when $\mathbf{v}^s_{\mathbf{k}_0}=0$ or $\mathbf{v}^{s'}_{\mathbf{k}_0+\bf{p}}=0$, or $\mathbf{v}^s_{\mathbf{k}_0} \parallel \mathbf{v}^{s'}_{\mathbf{k}_0+\bf{p}}$. The latter, for example, explains backscattering in metals where quasiparticle of $\bk$ is scattered to $-\bk$. The condition of singularity also could be viewed as an envelope curve for one-parameter $\mathbf{p}$ curve family  $E^{s'}_{\mathbf{k}+\bf{p}}=\omega$ and $E^s_{\mathbf{k}}=\omega$. This allows us to directly draw QPI pattern based on contours of constant energy.  It should be pointed out that the condition Eq.~\eqref{jacobi} can be related to the stationary phase approximation~\cite{liu_12}, which has been applied in the study of QPI for surface state of $\text{Bi}_2\text{Te}_3$.

The expression of Eq.~\eqref{JDOS_exact}, combined with geometrical representation of $F^{ss'}_{\alpha\beta}$ as discussed in Sec.~\eqref{sec:analysis_F}, allows an analysis of QPI pattern from the band structure directly. In the following, we will first give examples for simple models of gapless topological systems.

\subsection{Global patterns in QPI}
For a simple circle-like CCE, singularities come from backscattering from $\bk_0$ to $-\bk_0$. The topology of Dirac points manifests in their structures of spin-momentum binding, and will have effects on QPI patterns through the factor $F^{ss'}_{\alpha\beta}$ that may suppress the singularities.  For illustration, we choose the Hamiltonian,
\begin{equation}\label{ham:ideal-dirac}
\mathcal{H}_{\bk}=k^n(\cos{n\theta}~ \sigma_1+\sin{n\theta}~\sigma_2)
\end{equation}
where $n$ is an arbitrary integer, as a model for a Dirac point with topological charge/number $Q=n$~\cite{zhao_13,min_08}. Those systems have circle-like CCEs. Quasiparticle with eigenenergy $\pm|\bk|^n$ has the wavefunction $u_{\bk,\pm}\text{exp}(i\bk\cdot\mathbf{r})$, where $u_{\bk,\pm}=(1,\pm e^{-in\theta})^T$. We chose positive $\omega$, and indexes $s$ and $s'$ in the RRF framework are omitted.

\begin{figure}[ht]
	\includegraphics[width=0.8\columnwidth]{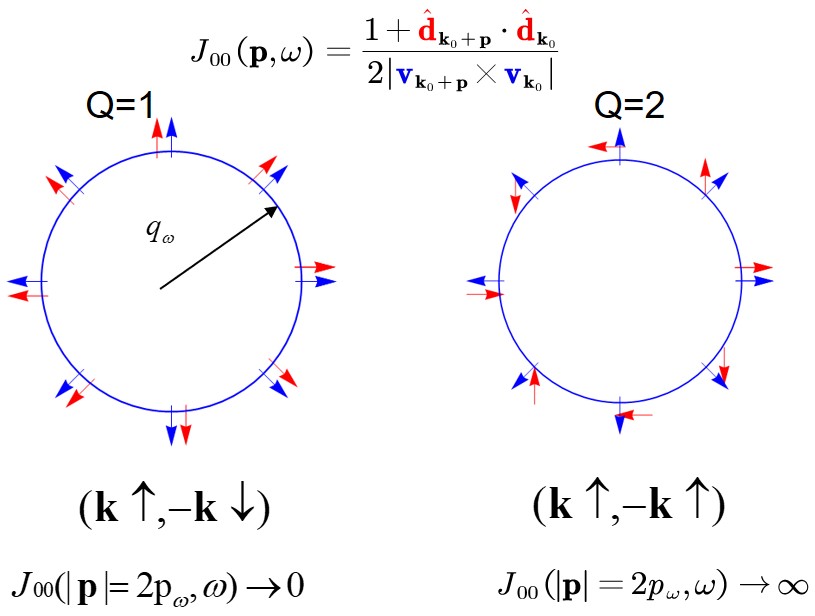}
	\caption{Illustration of how spin direction(red arrows) and group velocity(blue arrows) along the CCE jointly determine QPI pattern. For backscattering group velocities are reversed and the denominator becomes divergent. However, spin directions are reversed for $Q=1$ but are the same for $Q=2$. Thus the divergence is kept only for $Q=2$.    }
	\label{fig:jdos_00}
\end{figure}

We first consider the charge scattering and charge probe channel ($\alpha=\beta=0$) for systems with different topological numbers $Q$. In Fig.~\eqref{fig:jdos_00} we demonstrate how to use GJDOS of Eq.~\eqref{JDOS_exact} to analyze QPI patterns. A distinct behavior  can be identified between $Q=1$ and $Q=2$ that origins from their distinct structures of spin-momentum locking: QPI patterns are suppressed for $Q=1$ while restored for $Q=2$ for backscattering.

\begin{figure}
	\centering
	\includegraphics[width=0.8\columnwidth]{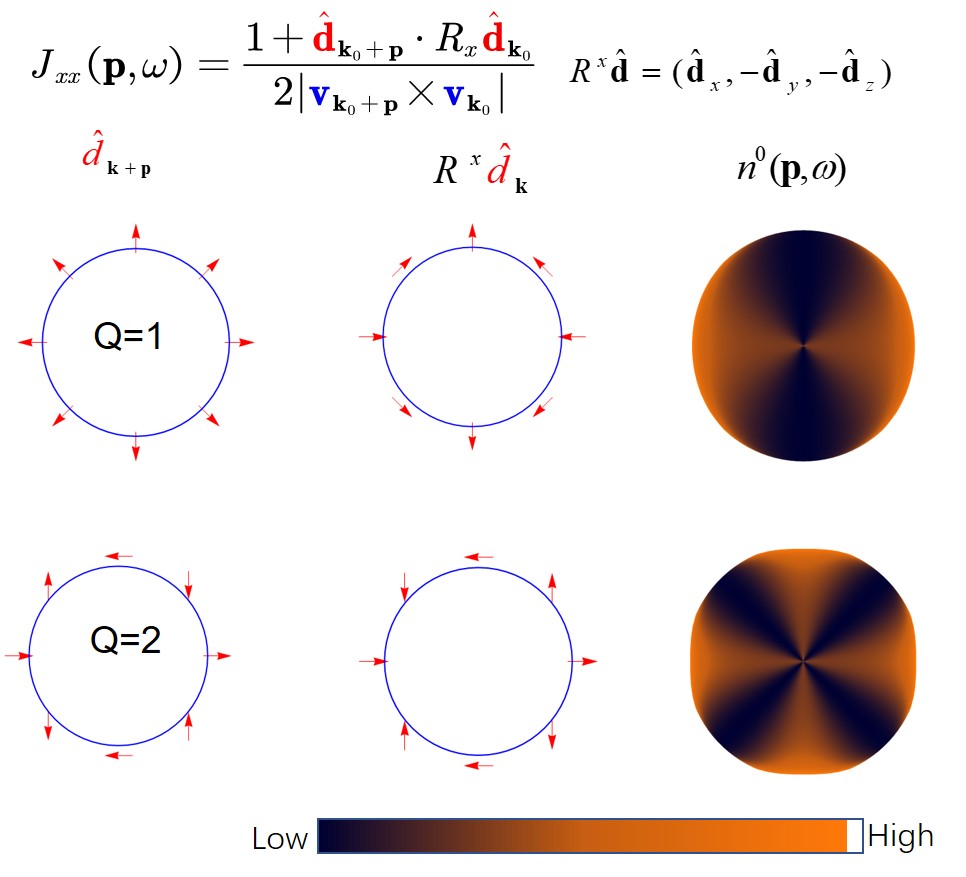}
	\caption{Illustration of effective direction-selective prohibition of backscattering. The combination of scattering scattering channel $\alpha=x$ and probe channel $\beta=x$ effectively rotates the spin of quasiparticle around $x$ axis by $\pi$, leading to a twist of spin interference, and backscattering is prohibited along specified directions. For $Q=1$ and $Q=2$ QPI patterns have two and four hot arcs (brown color), respectively.}
	\label{fig:jdos_xx}
\end{figure}

We now consider $\alpha=\beta$ channel. Note that $F_{\alpha\alpha}=2(1+\hat{\bd}_\bkp\cdot\hat R_\alpha\hat{\bd}_\bk)$. As the spin lies on $x-y$ plane, $R_z$ will not change $\hat{\bd}_\bk$. Interestingly, $F_{zz}$ is the same as that of $F_{00}$. Thus, although charge probe channel can not detect magnetic impurity $\sigma_z$ (the well-known prohibition of backscattering for Dirac fermion), a spin $\sigma_z$ resolved probe can~\cite{guo_10}. On the other hand, $R_{x/y}$ will change the sign of $\hat{d}^{y/x}_\bk$. As a result, for channel $\alpha=\beta=x/y$ suppressions of singularities will become direct dependent. In Fig.~\eqref{fig:jdos_xx} we show the case $\alpha=\beta=x$ for both $Q=1$ and $Q=2$, and QPI patterns have two and four separated hot arcs, separately. One can derive that the number of hot arcs turns to be $2n$ if the topological number $Q=n$, making it a global QPI pattern that can reveal topological number of the underlying system.

Following the above examples we can exploit global QPI patterns from different combinations of scattering and probe channels as indicators for topological numbers of the underlying systems. We give some topological-number indicators as following:
\begin{enumerate}
	\item \emph{Odd-even indicator}.  For scattering channel $\alpha=0$ and probe channel $\beta=0$, the spin coherent factor reads as $F_{00}\sim(1+\cos n(\theta_{-\bk_0}-\theta_{\bk_0}))$.  As $\theta_{-\bk_0}-\theta_{\bk_0}=\pi$, then $\mathcal{J}_{00}\sim \lim\limits_{\theta\ra\pi}\frac{F_{00}}{\sin\theta}\sim 0$ is zero only for odd $n$. This can be understood from the picture of spin-momentum locking, spins would be antiparallel/parallel for Dirac point with odd/even topological number at a pair $(-\bk_0,\bk_0)$. Thus, circle-like QPI patterns remain/disappear for even/odd topological numbers.
    \item \emph{Integer indicator}. For scattering channel $\alpha=x$ and probe channel $\beta=x$, the spin coherent factor reads as $F_{xx}\sim(1+\cos n(2\theta_{\bk_0}+\pi))$. There is a $2n$-fold periodicity with angle $\theta$. Consequently, there are $2|n|$ disconnected hot arcs in the QPI pattern. For $\alpha=\beta=y$, the conclusion holds while hot arcs rotate $\frac{\pi}{2n}$ compared to the case $\alpha=\beta=x$.
    \item\emph{Positive-negative indicator.} Counting hot arcs alone can not tell whether topological charge is positive or negative, thus we need an additional indicator. This can be achieved by rotating the scattering channel and probe channel a little, e.g., rotating clockwise, and observing how QPI pattern would rotate. If it rotates clockwise~(anti-clockwise), then topological charge is positive~(negative).
\end{enumerate}

Further, we can explore quasiparticle interference of two different Dirac points. For instance, intervalley scattering in graphene where two valleys have opposite topological numbers~\cite{mallet_12}. Consider two Dirac points with topological charge $Q_1=n$ and $Q_2=m$, following the analysis of $F_{\alpha\beta}$, we find that the odd-even indicator($\alpha=\beta=0$) would tell odd or even for $n-m$, and integers indicator would give $n+m$ disconnected hot arcs. Those results  provide  further clues to identify topological charges of different Dirac points, which would be demonstrated on QPI patterns of the graphene family.

\section{Applications to topological matters}
\label{sec:qpi-materials}
We now apply the reduced response function approach for topological materials~\cite{hasan_10,qi_11,zhao_13}. We chose three representative examples, including surface states of topological insulators $\text{Bi}_2\text{Te}_3$~\cite{fu_09,zhang_09} and BiTeI~\cite{crepaldi_12,kohsaka_15}, as well as the graphene family~\cite{geim_07}. For those real materials there may be deviations of  both dispersion and spin-momentum locking from ideal topological Dirac points described by ${\mathcal{H}}_n(\bk)$ (Eq.~\eqref{ham:ideal-dirac}). Thus, they provides playgrounds for studying both local and global patterns of quasiparticle interference.


\subsection{Surface states of topological insulator: $\text{Bi}_2\text{Te}_3$}
\label{subsection:qpi-ti}
In the surface state of $\text{Bi}_2\text{Te}_3$, wrapping terms appears~\cite{fu_09}.  At the low energy limit, the CCE can be approximated as a circle, while at larger energy it becomes non-convex. Such distinct geometries of CCEs will lead to different patterns in quasiparticles interference at different energies~\cite{zhou_09,lee_09_TI}. On one hand, we expect remarkable new features in QPI; on the other hand, we can testify whether indicators of topological number are robust under complex CCEs when the energy increases.

To study the effect risen up by the complicated geometry of CCE at large energy,
we consider the surface state of topological insulator $\text{Bi}_2\text{Te}_3$ which can be modeled as~\cite{fu_09,lee_09_TI},
\begin{equation}\label{ham:BiTe}
{\mathcal{H}}_{TI}(\bk)=v(k_x\sigma_y-k_y\sigma_x)+\lambda k^3\cos{3\theta_\bk}\sigma_z,
\end{equation}
where $\theta_\bk=\arctan{k_y/k_x}$. The system ${\mathcal{H}}_{TI}(\bk)$ has time-reversal symmetry and belongs to AII class. It owns a $Z_2$ type topological number of $Q=1$~\cite{zhao_13}. Dispersions  of two branches $s=\pm1$ are opposite and we take only the positive one ($s=1$) for consideration in RRF. At a low energy, the CCE is circle-like, and QPI patterns under different combinations of scattering and probe channels are consistent with those of ideal Dirac point with $Q=1$ (see Fig.~(6) for details).
\begin{figure}
	\centering
	\includegraphics[width=1.0\columnwidth]{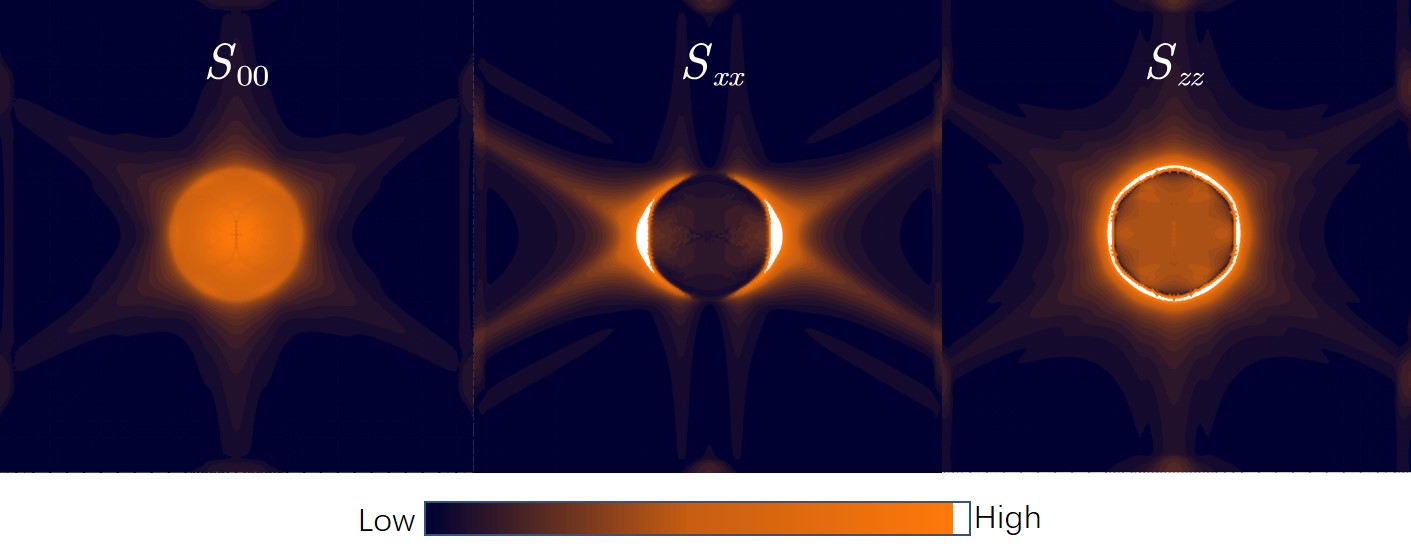}
	\caption{QPI patterns for surface state of 3D topological insulator $\text{Bi}_2\text{Te}_3$ for circle-like contour of constant of energy at low energy $\omega=0.25$. Backscattering is prohibited for charge impurity and charge probe, as evidenced $\mathcal{S}_{00}$, while is restored for $\sigma_z$ impurity and $\sigma_z$-resolved probe.  The QPI pattern shows two bright arcs for $\sigma_x$ impurity and $\sigma_x$-resolved probe. Those QPI patterns under different channels reveal a topological number $Q=1$.}
	\label{fig:BiTe-QPI-small}
\end{figure}

\begin{figure}
	\centering
	\includegraphics[width=1.0\columnwidth]{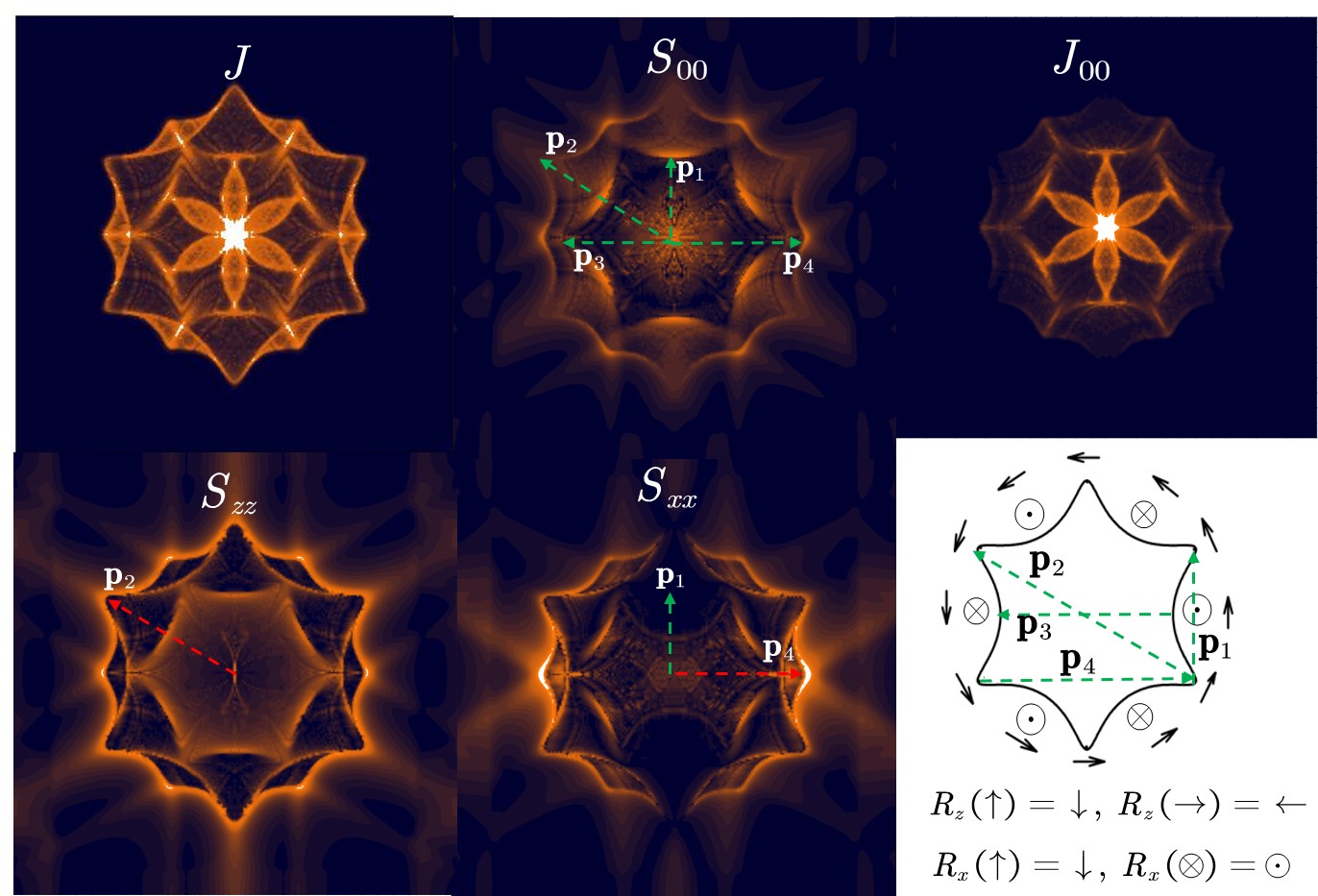}
	\caption{QPI patterns for non-convex contour of constant energy for surface state of 3D topological insulator $\text{Bi}_2\text{Te}_3$. CCE at $\omega=1$ is given in the right-bottom (parameters in Eq.~\eqref{ham:BiTe} are $v=1$ and $\lambda=2$). Joint density of states $J$ evidences high degrees of singularities with hotter spots, e.g., $\bp_1,\bp_4$ due to zero joint curvatures, and $\bp_3$ due to near-nesting. QPI patterns are given by $\mathcal{S}_{\alpha\alpha}$ for $\alpha=0,x,z$ scattering and probe channels.  As a comparison with $\mathcal{S}_{00}$, $\mathcal{J}_{00}$ gives extra false features around the center.}
	\label{fig:BeTe-RRF}
\end{figure}

For a sufficient large energy, the wrapping term becomes important and CCE turns to be nonconvex and exhibit six cusps. A scattering between a pair of cusps can satisfy the condition of singularity Eq.~\eqref{jacobi} but is not backscattering-type. Remarkably, $\bp_1$ and $\bp_4$ correspond to zero joint curvature.
Moreover, $\bp_3$ is a backscattering scattering between near-nesting arcs of CCE. Those vectors ($\bp_1,\bp_3,\bp_4$) thus have higher degrees of singularity than $\frac{1}{2}$, evidenced by hotter spots than $\bp_4$ in joint density of states $J(\bp,\omega)$ (See Fig.~\eqref{fig:BeTe-RRF}).  QPI under different combinations of scattering and probe channels exhibits more complex patterns as a consequence of both topological and geometrical aspects of the band structure.  Numeral evaluation of $S_{00}(\bp,\omega)$ shows hot spots locating at  $\bp_1, \bp_4$ that are absent for small energy when CCE is convex.  Due to the spin coherent factor, spot is darker for $\bp_4$ than that of $\bp_1$. Those results fit very well with the FT-STS experiment~\cite{roushan_09,beidenkopf_11}. We also present $J_{00}(\bp,\omega)$ as a comparison. It can be seen that $J_{00}(\bp,\omega)$ has some sharp features around the center. These features come from joint densities of states with the same arc of the CCE, and are not stable singularities. Thus, a direct application of JDOS should take those as false features, when QPI should be given by the real part of the RRF.

We then investigate QPI in the other situations, e.g, $S_{xx}(\bp,\omega)$ and $S_{zz}(\bp,\omega)$. Compared with QPI at low energy, we can see that while local features
are different, the global patterns persist. For $S_{zz}(\bp,\omega)$, the bright closed curve results from backscattering. For $S_{xx}(\bp,\omega)$, QPI pattern is cut into two parts in the middle, showing a 2-fold pattern. Remarkably, $\bp_4$ is enhanced while $\bp_1$ is suppressed, due to the effective $R_x$ reflection of the spin of scattered-in quasiparticle. These distortions of local features plus persistences of global patterns suggest the robustness of indicators for topological numbers.  To reveal these QPI patterns, magnetic impurity with specified direction  as well as spin-resolved STS techniques should be required, which are expected to be fulfilled in future FT-STS experiments~\cite{wiesendanger_09,jeon_17,cornils_17}.

\subsection{Surface state of polar semiconductor: BiTeI}
\label{subsection:BiTeI}
\begin{figure}
	\centering
	\includegraphics[width=0.8\columnwidth]{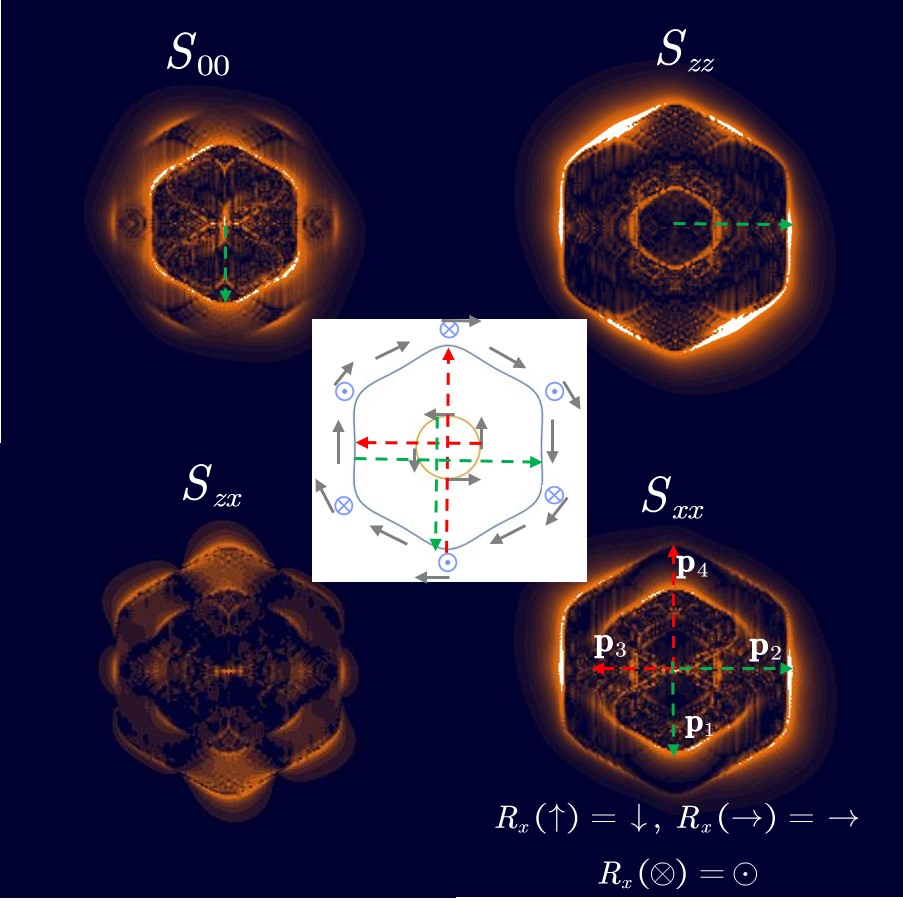}
	\caption{Quasiparticle interference for the surface state of BiTeI. In the middle is an inset of contours of constant energy at $\omega=-0.01$, which has two concentric CCEs. QPI for different combinations of scattering and probing channels are given by $\mathcal{S}_{00},\mathcal{S}_{zz},\mathcal{S}_{zx},\mathcal{S}_{xx}$, respectively. Inter-CCE (intra-CCE) scattering are favored (prohibited) in $\mathcal{S}_{00}$, while  Intra-CCE (inter-CCE) scattering are favored (prohibited) in $\mathcal{S}_{zz}$.  A global pattern of 2-fold symmetry appear in both $\mathcal{S}_{zx}$ and $\mathcal{S}_{xx}$. The four hot arcs in $\mathcal{S}_{xx}$ can be explained by direction selective spin coherent factor due to an effective spin reflection around the $x$ direction $\hat{R}_x$. }
	\label{fig:BiTeI-RRF}
\end{figure}

We continue to study another topological state: surface state of polar semiconductor BiTeI~\cite{crepaldi_12,kohsaka_15,kohsaka_17}. There are two concentric CCEs, due to a contribution from the bulk state. Those two CCEs correspond to two branches of $s=1,-1$, and thus are opposite in  spin-momentum locking along the CCE (see Fig.~\eqref{fig:BiTeI-RRF}). Scattering between two concentric CCEs can not be ignored.  Moreover, the outer CCE has a larger distortion from the circle.  Such a system thus exhibits unconventional properties of the band structure that can lead to novel patterns of quasiparticle interference.

Following Ref.~\cite{kohsaka_15}, the surface state can be modeled as
\begin{align}
H_0(k_x,k_y) &=\left(E_0 + \frac{k^2}{2m}E(k)\right)I + V(k)(k_x\sigma_y-k_y\sigma_x)
\nonumber\\ &+\Lambda(k)(3k_x^2-k_y^2)k_y\sigma_z,
\label{ham:BiTeI}
\end{align}
where $I$ is the identity matrix, 
and $k=\sqrt{k_x^2+k_y^2}$. The last term of Eq.~\eqref{ham:BiTeI} respects the C$_\mathrm{3v}$ symmetry of BiTeI~\cite{ishizaka_11,crepaldi_12}. Functions are $E(k)=1+\alpha_4k^2 + \alpha_6k^4$, $V(k)=v(1+\beta_3k^2+\beta_5k^4)$ and $\Lambda(k)=\lambda(1+\gamma_5k^2)$.  Parameters as set as
$m = 0.0168$~eV$^{-1}$\AA$^{-2}$,
$\alpha_4 = -2.03$~\AA$^{-2}$,
$\alpha_6 = 87.5$~\AA$^{-4}$,
$v = 3.13$~eV\AA$^{-1}$,
$\beta_3 = -2.01$~\AA$^{-2}$,
$\beta_5 = 323$~\AA$^{-4}$,
$\lambda = -41.7$~eV\AA$^{-3}$,
$\gamma_5 = 2.43$~\AA$^{-2}$, and
$E_0 = -0.352$~eV.

In the treatment of RRF, we should take all inter and intra CCEs scattering by a summation over $s=\pm1$ and $s'=\pm1$ in Eq.~\eqref{eq:cal_RF}. Numeral simulations are given in Fig.~\eqref{fig:BiTeI-RRF} for $\mathcal{S}_{00},\mathcal{S}_{zz},\mathcal{S}_{zx},\mathcal{S}_{xx}$. As we can see in $\mathcal{S}_{00}$, intra-CCE backscattering is prohibited for both the inner and the outer CCEs, due to a reversal of spin directions. However, inter-CCE scattering is allowed
and a bright hexagon appears. QPI patterns in the case of $\mathcal{S}_{zz}$ is reversed: bright hexagon appears only for intra-CCE scattering. The most interesting case is spin $\sigma_x$ scattering and $\sigma_x$ probing channels: there are four disconnected bright arcs that respect a 2-fold symmetry.  The outer arcs come from intra-CCE scattering: since $\hat{R}_x$ reflection will not change spin in $x$ direction, quasiparticle interference for $\bp_3$ is still suppressed due to an approximately reversal of spins. However, hot arcs arises for wavevector $\bp_2$, as $\hat{R}_x(\uparrow)=\downarrow$, thus the spin coherent factor now becomes nonzero. The inner arcs result from inter-CCE scattering: quasiparticle interference of wave vector $\bp_1$ is enhanced while $\bp_3$ is suppressed. The outer and inter hot arcs locate at horizon and vertical directions, respectively. Thus, they still respect the 2-fold symmetry. We also show $\mathcal{S}_{zx}$ that also respect the 2-fold symmetry.  We remark that the $2$-fold symmetry can be considered as global patterns for indicting topological numbers $Q=1$, instead of merely counting the number of disconnected hot arcs.

\subsection{The graphene family}
Graphene and its relatives of N-layer graphene host two topological Dirac points with opposite topological numbers~\cite{min_08}. Moreover, the topological number can vary with the layer.  For example, there are two gapless points with $Q=\pm 2$ for AB stacked bilayer graphene and $Q=\pm 3$ for ABC stacked trilayer graphene. Those make the graphene family an ideal platform for studying global patterns of quasiparticle interferences.

We consider N-layer graphene with ABC stacking. For the i-th layer two sublattices are denoted as $a_i$ and $b_i$, then there are isolated atoms $a_1$ and $b_N$ that are not directly coupled to other layers. Taking $a_1$ and $b_N$ as a pseudospin, we have an effective Hamiltonian $H_{G,N}(\bk)=(\Re{f(\bk)})^N\sigma_x+(\Im{f(\bk)})^N\sigma_y$ to model ABC stacking N-layer graphene~\cite{min_08}. Here $f(\bk)=1+e^{i\bk\cdot\bf{G}_1}+e^{i\bk\cdot\bf{G}_2}$ with $\bf{G}_1,\bf{G}_2$ being primitive vectors for graphene. Topological Dirac points locate at  $\mathbf{K}$ and $\mathbf{K}'$ with topological charges $Q=\pm N$ respectively. At low energy limit $H_{G,N}(\bk)$ can be approximated by $H_{\pm N}(\bk)$ with $f(k) \sim k^N$.  It should be noted that the pseudospin here  corresponds to sublattice, and the physical meaning of scattering and probing channel should be adjusted accordingly.  For  point-like impurity locating at site $a/b$ the scattering channel is $\tau_{a/b}=\frac{1}{2}(\sigma_0\pm\sigma_z)$ (here we have shortnoted $a/b$ for $a_1/b_N$). Similarly, LDOS is measured at site $a/b$ and probing channel is $\tau_{a/b}$.  The reduced response functions $\mathcal{R}_{00}(\bp,\omega)$ and
$\mathcal{R}_{zz}(\bp,\omega)$ can be obtained  as follows,
$\mathcal{R}_{00}(\bp,\omega)=\frac{1}{2}(\mathcal{R}_{aa}(\bp,\omega)+\mathcal{R}_{bb}(\bp,\omega)+\mathcal{R}_{ab}(\bp,\omega)+\mathcal{R}_{ba}(\bp,\omega)+)$,$ \mathcal{R}_{zz}(\bp,\omega)=\frac{1}{2}(\mathcal{R}_{aa}(\bp,\omega)+\mathcal{R}_{bb}(\bp,\omega)-\mathcal{R}_{ab}(\bp,\omega)-\mathcal{R}_{ba}(\bp,\omega))$,
where $\mathcal{R}_{\mu\nu}$($\mu=a,b$ and $\nu=a,b$) correspond to probe channel $\tau_\mu$ and scattering channel $\tau_\nu$.
\begin{figure}
	\centering
	\includegraphics[width=1.0\columnwidth]{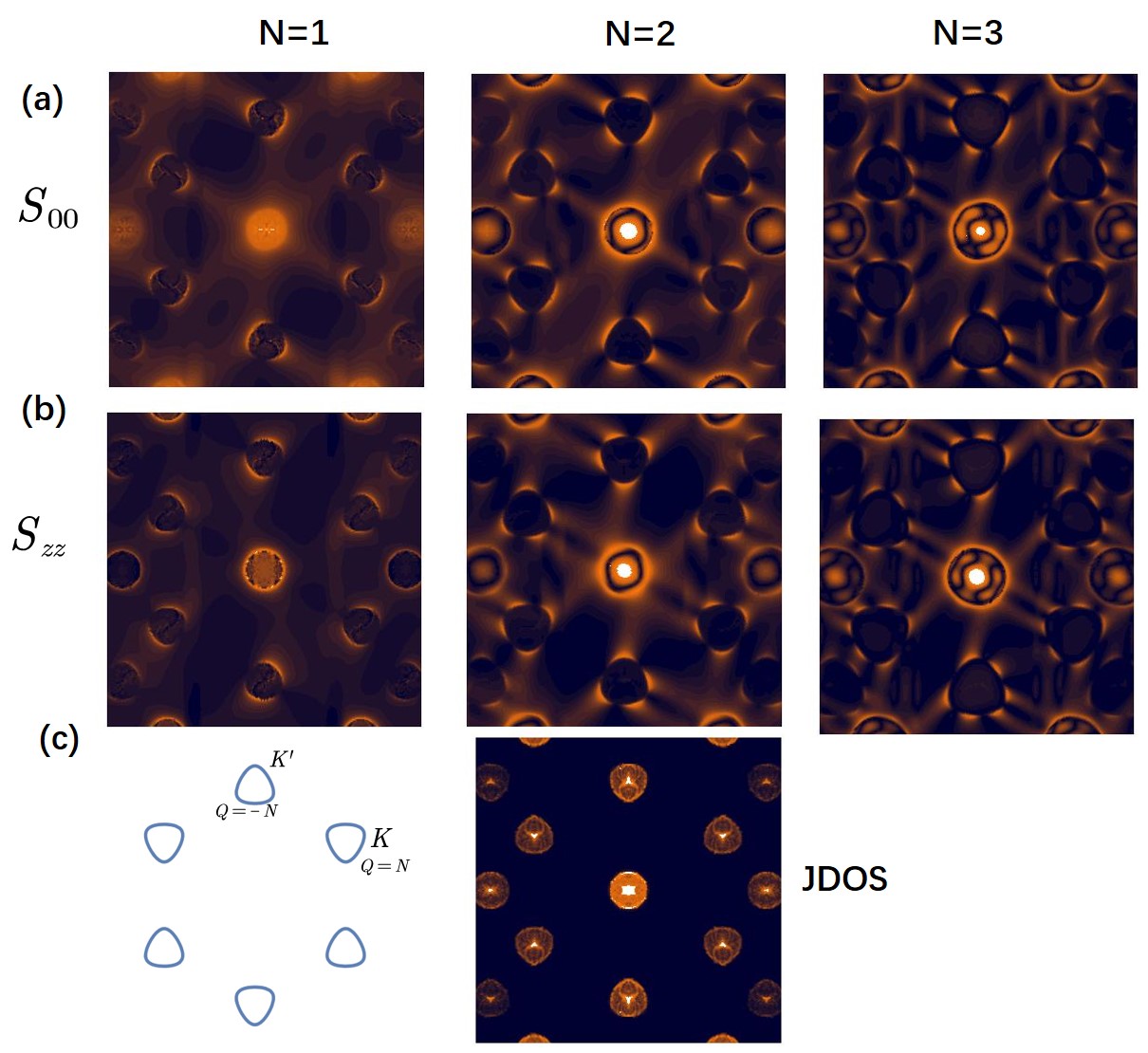}
	\caption{QPI pattern for graphene family evaluated by  $\mathcal{S}_{00}(\bp,\omega)$ and $\mathcal{S}_{zz}(\bp,\omega)$, as shown in (a) and (b), respectively. Here we chose $\omega=0.3,0.15,0.1$ for $N=1,2,3$, respectively. Inter-valley scattering leads to remarkable $2N$ hot arcs for $N=1,2,3$. CCEs and the joint density of states are shown in (c) for $N=1$. }
	\label{fig:qpi-graphene}
\end{figure}

As two valleys $\bf{K}$ and $\bf{K'}$ have the opposite topological charges, it is expected to see distinct QPI patterns between the center of $\bp$-space due to intravalley scattering and those due to intervalley scattering away from the centre, such as around $\pm 2\mathbf{K}$, $\pm 2\mathbf{K}'$ and $\pm 2(\mathbf{K}-\mathbf{K}')$. We numeral evaluate  $\mathbf{S}_{00}(\bp,\omega)$ and $\mathbf{S}_{zz}(\bp,\omega)$ for ABC stacked N-layer graphene for $N=1, 2, 3$ respectively, as seen in Fig.~\eqref{fig:qpi-graphene}.  For both channels, distinct $2N$ pieces of disconnected hot arcs appear for intravalley scattering and are rotated relatively. Those distinct features of QPI both for intervalley scattering indicate opposite topological charges for Dirac point located at the valleys. The simulations fit well with the FT-STS experiments~\cite{brihuega_08,mallet_12} for monolayer graphene, while  better resolution is required for bilayer graphene.  For intravalley scattering, a bright circle appears for channels $\alpha=\beta=z$ and is absent $\alpha=\beta=0$ for monolayer graphene. The persistent presence of bright circle around the center for bilayer and trilayer graphene may result from high density of states from dispersions of $k^N$
($N>1$). We mention that joint density of states give false hot spots at $\bp=0,\pm 2\mathbf{K},\pm 2\mathbf{K}$, and $\pm 2(\mathbf{K}-\mathbf{K}')$.

\section{Conclusions}
We have developed a reduced response function approach for analyzing and simulating quasiparticle interference under different scattering and probing channels. The  applicability of the generalized joint density of state has been clarified. We have analytically shown how singularities and spin coherent factor jointly determine QPI patterns. Remarkably, those local features and global patterns in QPI can be used to infer geometry details of dispersions and topology of band structures for the underlying system. We have also proposed indicators of topological numbers from global patterns of QPI for topological Dirac points. We have numerally simulated QPI of different topological materials, whose complicated geometrical features yet robust global patterns are evidenced and can be nicely captured in the RRF framework.

We remark that advances in spin-resolved STS technique~\cite{wiesendanger_09,jeon_17,cornils_17} may experimentally visualize novel QPI patterns explored in this paper. Further investigations along this line include finding QPI that should be explained by the generalized joint density states, and an extension of the reduced response function approach for systems beyond two-band Hamiltonian description.

\section{acknowledgement}
The work was supported by the GRF (No. HKU173057/17P) and CRF (No. C6005-17G) of Hong Kong, and the NKRDP of China (Grant No. 2016YFA0301800).

%

\end{document}